\documentclass[a4paper,11pt]{article}
\usepackage{epsfig,amsbsy,array,multirow}

\newcommand{\bsm}{\begin{small}} 
\newcommand{\esm}{\end{small}} 
\newcommand{\bc}{\begin{center}} 
\newcommand{\ec}{\end{center}} 
\newcommand{\bl}{\begin{large}} 
\newcommand{\el}{\end{large}}
\newcommand{\bL}{\begin{Large}} 
\newcommand{\eL}{\end{Large}} 
\newcommand{\bh}{\begin{huge}} 
\newcommand{\eh}{\end{huge}} 
\newcommand{\ben}{\begin{enumerate}} 
\newcommand{\een}{\end{enumerate}} 
\newcommand{\bit}{\begin{itemize}} 
\newcommand{\eit}{\end{itemize}} 
\newcommand{\bq}{\begin{equation}} 
\newcommand{\eq}{\end{equation}} 
\newcommand{\bqa}{\begin{eqnarray}} 
\newcommand{\eqa}{\end{eqnarray}} 
\newcommand{\nn}{\nonumber}

\newcommand{\nl}{\nonumber \\}
\newcommand{\suml}{\sum\limits}
\newcommand{\intl}{\int\limits}

\def\demo{$\Delta\eta\mu \acute{o} \kappa \varrho \iota \tau o \varsigma$}

\begin{document}

\title{ {\bf  Parton counting: physical and computational complexity of multi-jet production
at hadron colliders} }

\author{Petros D. Draggiotis\thanks{petros@sci.kun.nl, 
				    $^\dagger$kleiss@sci.kun.nl}~ \\
         {\it University of Nijmegen, Nijmegen, the Netherlands} \\
         and \\
         {\it Institute of Nuclear Physics, NCSR \demo, 15310 Athens, Greece} \\
\vspace*{0.2cm} \\
	Ronald Kleiss$^\dagger$\\        
        {\it University of Nijmegen, Nijmegen, the Netherlands}}

\maketitle

\begin{abstract}
We present an enumeration of all possible amplitudes that contribute to an 
$n$-jet process in QCD. We estimate the number of amplitudes for large
number of jets and determine the actual number of amplitudes to be
calculated, which is smaller due to relabelling among (massless) quark flavours.
\end{abstract}

\section{Introduction}
With the advent of high-energy hadron colliders such as the Tevatron and
the LHC, there arises a need for accurate QCD calculations of amplitudes of
increasing complexity, either as problems in their own right,
or as possible backgrounds to other physics. The complexity
becomes apparent either in the number of loops to be considered,
or in the number of external legs in the diagrams: the present paper aims to
deal with the latter of these issues.
In recent years there has been considerable progress in the computation
of multi-leg QCD amplitudes \cite{CarMor,DKP,Mangano}. Essentially based on
the earlier work of \cite{BerGie,Kuijff}, these new algorithms employ the
recusrive structure of the Schwinger-Dyson equations to express the full
amplitude in terms of smaller subamplitudes in the essentially most compact
way, thereby reducing the computational complexity of these amplitudes
from roughly $k!$ to about $3^k$, where $k$ is the number of external legs.
This enormous improvement has led to the becoming feasible of
amplitudes with as many as 8 or 9 outgoing partons.

Another issue then arises, that of summing the contributions of all
possible QCD processes to a given multi-jet final state. As we shall show, the
number of QCD amplitudes contributing to the probability of an observed event
increases very rapidly with the number of jets, so that the following
questions become relevant. How many amplitudes contribute precisely?
What is the asymptotic form of this number for large multiplicity?
To what extent is the folk-lore that the purely gluonic amplitude dominates
the cross section still valid? Given that many amplitudes can be
related to each other by a simple relabelling of the quark flavours, 
how many {\em distinct\/} amplitudes have to
be computed? 
These are the questions addressed in this paper.

\section{Physical complexity: contributing amplitudes}
Under the
assumption that
the various (anti)quark types and gluons cannot be distinguished
experimentally, and that all these parton types are (essentially) massless,
the only information experimentally available about any given event is
the configuration of the observed momenta. We shall denote such an
event by its momenta as follows:
$$
p_1 + p_2 \;\;\rightarrow\;\;
q_1 + q_2 + q_3 + \cdots + q_n\;\;\;,
$$
where $n$ outgoing partons/jets are observed. To obtain the total probability
density for this event in phase space, one has to consider all possible
$2\to n$ QCD amplitudes\footnote{Of course, squared and spin/colour-summed
and -averaged in the usual way.},viewed here as functions of $2+n$
momentum arguments, and assign the observed momenta to these arguments in all
possible ways without double counting. Note that, due to the composite nature
of the incoming hadrons, also the initial state may require more than
one assignment: for instance, a quark-gluon initial state $q(p_1)g(p_2)$
is to be counted as distinct from $q(p_2)g(p_1)$, whereas of course the
purely gluonic initial state $g(p_1)g(p_2)$ is counted only once. Similarly,
if the final states contains $m$ quarks of a certain type, a corresponding
factor $1/m!$ has to be applied. In what follows we shall denote the
number of (essentially) massless quarks in the final state by $f$,
so that $f=4$ at relatively low momenta where the $b$ quark might be
identifiable, $f=5$ typically for QCD studies at the LHC, and $f$ would be
6 at some future multi-hundred TeV collider. The number of flavours
contributing appreciably in the initial state is denoted by $j$, so that
$j=3$ would be appropriate if the charm quark structure function can be
neglected, and $j=4$ if it is included. We shall, however, keep to general
$j$ and $f$ as much as possible.

\subsection{Arrangement of the initial states and final states}
The various possibilities for the initial states are:
\begin{itemize}
 \item $\boldsymbol{gg}$ Obviously, there is only one possibility
       for the initial state. The final state can be anything so
       there are $n$ partons in the final state.
 \item $\boldsymbol{q_i \bar{q}_i}$ If we have $j$ flavours then
       there are $2j$ possibilities for the initial state and $n$
       partons to be distributed in the final state.
 \item $\boldsymbol{g q_i  \, ,g \bar{q}_i }$ There are $2j$ initial
       states as in the previous case. For the final state, we know
       that there must at least one $q_i$ or $\bar{q}_i$ so there are
       $(n-1) + q_i$ or $(n-1) + \bar{q}_i$  partons.
 \item $\boldsymbol{q_i q_i  \, ,\bar{q}_i \bar{q}_i }$ For the scattering
       of identical quarks (anti-quarks), there are $j$ initial states and
       since the same partons must appear in the final state we can have
       $n-2$ partons plus the initial quarks or anti-quarks.
 \item $\boldsymbol{q_i q_k  \, ,\bar{q}_i \bar{q}_k  \, ,i \neq k }$
       For the scattering of different quarks (anti-quarks) we have
       $j(j-1)$ possibilities for the initial state, and again
       $n-2$ partons plus the initial quarks (anti-quarks), for the
       final state.
 \item $\boldsymbol{q_i \bar{q}_k  \, ,i \neq k }$ For this final case 
       we have $2j(j-1)$ initial states and $n-2$ partons plus the quark and
       the anti-quark in the final state.
\end{itemize}
All of the above can be summarized in the following table.

\begin{center}
\begin{tabular}{|c|c|r|}
\hline
Initial States & \# possibilities & Final States \\
\hline
\hline
$ gg $ & $1$ & $n$ \\
\hline
$ q_i \bar{q}_i$ & $2j$ & $n$ \\
\hline
$ gq_i $ & $2j$ & $(n-1)+q_i$ \\
\hline
$ g\bar{q}_i $ & $2j$ & $(n-1)+\bar{q}_i$ \\
\hline
$ q_i q_i $ & $j$ & $(n-2)+q_i+q_i$ \\
\hline
$ q_i q_j, i \neq j $ & $j(j-1)$ & $(n-2)+q_i+q_j$ \\
\hline
$ \bar{q}_i \bar{q}_i $ & $j$ & $(n-2)+\bar{q}_i+\bar{q}_i$ \\
\hline
$ \bar{q}_i \bar{q}_j, i \neq j $ & $j(j-1)$ & $(n-2)+\bar{q}_i+\bar{q}_j$ \\
\hline
$ q_i \bar{q}_j, i \neq j $ & $2j(j-1)$ & $(n-2)+q_i+\bar{q}_j$ \\
\hline
\end{tabular}
\end{center}
where $i,k=1,\ldots,j$. From the second column we can read off the total number 
of initial-state momentum configurations:
\bq
1+3(2j)+j+j(j-1)+j+j(j-1)+2j(j-1)=(1+2j)^2
\eq
From this table we can arrange four different groups of initial states, which
differ in the flavour structure of their final states. They are shown in the following table:

\begin{center}
\begin{tabular}{|c|c|c|}
\hline
Group & Initial state &\# of final states \\
\hline
\hline
$ A $ & $gg, \; q_i \bar{q}_i $ & $A(n)$ \\
\hline
$ B $ & $gq_i, \; g\bar{q}_i$ & $B(n)$ \\
\hline
$ C $ & $q_i q_i, \; \bar{q}_i \bar{q}_i$ & $C(n)$ \\
\hline
$ D $ & $q_i q_k, \; \bar{q}_i \bar{q}_k, \; q_i \bar{q}_k, i \neq k$ & $D(n)$ \\
\hline
\end{tabular}
\end{center}

\subsection{Counting of the final states}
\paragraph{Group A } 

The distinct possibilities for flavourless final states
are:
\bq
n=n_0 \ast (g)+ n_1 \ast (q_1\bar{q}_1) +n_2 \ast (q_2\bar{q}_2) +
\cdots + n_f \ast (q_f\bar{q}_f)
\eq
where $n_0$ is the number of gluons $g$ and $n_1,n_2, \ldots ,n_f$ are the numbers
of $q_f$ and  $\bar{q}_f$ quarks with different flavour $f$. 
The number of different processes $A(n)$ is the number of the various dinstinct ways 
to distribute $n$ different final momenta among $n$ partons:
\bq
A(n)=\sum_{n_0,n_1,\ldots,n_f \geq 0 } 
\frac{(n)!}{(n_0)!(n_1)!^2(n_2)!^2 \cdots (n_f)!^2}
\Theta(n_0+2n_1+2n_2+ \cdots +2n_f=n)
\eq
where $\Theta(a=b)=\delta_{a,b}$.
We can evaluate this number by forming the generating function: 
\bqa
\mathcal{A}(x)=\sum_{k \geq 0} \frac{x^{n}}{n!}A(n)&=& \sum_{n_0,n_1,\ldots,n_f \geq 0 }
\frac{x^{n_0}}{n_0!} \frac{x^{2n_1}}{(n_1)!^2}\frac{x^{2n_2}}{(n_2)!^2}
\cdots \frac{x^{2n_f}}{(n_f)!^2} \nn \\ 
&=& \left( \sum_{n \geq 0} \frac{x^{n}}{n!} \right) 
\left( \sum_{n \geq 0} \frac{x^{2n}}{(n)!^2} \right)^f 
= e^x \cdot I_0(2x)^f    
\eqa
where $I_0(x)$ is the modified Bessel function of the first kind and zeroth order \cite{Abra}.

\paragraph{Group B }

All the possible final states for this case have a single net flavour, and 
can be written as follows
\bq
n=n_0 \ast (g)+ n_1 \ast (q_1\bar{q}_1) +n_2 \ast (q_2\bar{q}_2) +
\cdots + n_f \ast (q_f\bar{q}_f)+q_i
\eq
The number of different processes $B(n)$ is:
\bq
B(n)=\sum_{n_0,n_1,\ldots,n_f \geq 0 }
\frac{(n_0+2n_1+2n_2+ \cdots +2n_f+1)!}{n_0!(n_1+1)!(n_1)!(n_2)!^2 \cdots (n_f)!^2}
\Theta(n_0+2n_1+2n_2+ \cdots +2n_f+1=n)
\eq
This gives the generating function
\bqa
\mathcal{B}(x)&=& \left( \sum_{n \geq 0} \frac{x^{n}}{n!} \right) 
\left( \sum_{n \geq 0} \frac{x^{2n}}{(n)!^2} \right)^{f-1}
\left( \sum_{n \geq 0} \frac{x^{2n+1}}{(n)!(n+1)!} \right) 
= e^x \cdot I_0(2x)^{f-1} \cdot  I^{\prime}_0(2x)    
\eqa
where the prime denotes the derivative of the Bessel function with respect
to the argument $2x$.

\paragraph{Group C }

For this case the final state is
\bq
n=2n_0 \ast (g)+ n_1 \ast (q_1\bar{q}_1) +n_2 \ast (q_2\bar{q}_2) +
\cdots + n_f \ast (q_f\bar{q}_f)+2 \ast (q_i)
\eq
and the number of possibilities is
\bq
C(n)=\sum_{n_0,n_1,\ldots,n_f \geq 0 }
\frac{(n_0+2n_1+2n_2+ \cdots +2n_f+2)!}{n_0!(n_1+2)!(n_1)!(n_2)!^2 \cdots (n_f)!^2}
\Theta(n_0+2n_1+2n_2+ \cdots +2n_f+2=n)
\eq
The generating function is
\bq
 \mathcal{C}(x)=\sum_{n \geq 0} \frac{x^{n}}{n!}C(n) = e^x I_0(2x)^{f-1} \cdot 
  \{ 2 I_0^{\prime \prime}(2x)-I_0(2x) 
\eq

\paragraph{Group D }

The derivation goes through as in the previous cases and the result is
\bq
\mathcal{D}(x)=\sum_{n \geq 0} \frac{x^{n}}{n!}D(n) = e^x I_0(2x)^{f-2} \cdot
 \left( I^{\prime}_0(2x) \right)^2 
\eq
The total number of possibilities for the final state can now be determined:
\bq
G(n)=(1+2j)A(n)+4jB(n)+2jC(n)+4j(j-1)D(n)
\eq
with the generating function
\bqa
\mathcal{G}(x)=\sum_{n \geq 0} \frac{x^{n}}{n!}G(n) = e^x 
&\{& (1+2j) \; I_0(2x)^{f} + 4j \; I_0(2x)^{f-1} \; I^{\prime}_0(2x) \nn \\
&+& 2j \; I_0(2x)^{f-1} \left( 2I^{\prime \prime}_0(2x) -  I_0(2x) \right) \nn \\
&+& 4j(j-1) \; I_0(2x)^{f-2} \; \left( I^{\prime}_0(2x) \right)^2 \; \; \} 
\label{Gen-Fun-alt}
\eqa
We can put this in a more compact form:
\bq
\mathcal{G}(x)=e^x I_0(2x)^{f-j} \left( 1+\frac{d}{dx} \right)^2 I_0(2x)^j
\label{Gen-Fun}
\eq
To get the number of processes we expand the generating function $\mathcal{G}(x)$ and pick
out the relevant coefficients. For example, for $f=3,4,5$ flavours we have:

\begin{center}
\begin{tabular}{|>{$}c<{$}||>{$}c<{$}|>{$}c<{$}||>{$}c<{$}|>{$}c<{$}|
                           |>{$}c<{$}|}
  \hline \multicolumn{6}{|>{$}c<{$}|}{\mathrm{Total \; number \; of \; amplitudes} }\\
  \hline \multicolumn{1}{|>{$}c<{$}||}{ }&
         \multicolumn{2}{>{$}c<{$}||}{f=3}&   
         \multicolumn{3}{>{$}c<{$}|}{f=4} \\   
  \hline n &j=2 &j=3 &j=2 &j=3 
                       &j=4 \\
  \hline 2 &71 &127 &81 &141 &217  \\
  \hline 3 &299 &511 &377 &625 &921  \\
  \hline 4 &1,763 &3,301 &2,645 &4,867 &7,761  \\
  \hline 5 &8,955 &16,297 &15,325 &27,087 &41,889  \\ 
  \hline 6 &54,353 &103,279 &113,733 &213,879 &345,465  \\
  \hline 7 &304,701 &570,367 &745,421 &1,364,811 &2,162,617  \\
  \hline 8 &1,879,723 &3,595,177 &5,704,061 &10,836,831 &17,605,249  
\\\hline
\end{tabular}
\end{center}

\begin{center}
\begin{tabular}{|>{$}c<{$}||>{$}c<{$}|>{$}c<{$}||>{$}c<{$}|>{$}c<{$}|}
  \hline \multicolumn{5}{|>{$}c<{$}|}{\mathrm{Total \; number \; of \; amplitudes }}\\
  \hline \multicolumn{1}{|>{$}c<{$}||}{ }&
         \multicolumn{4}{>{$}c<{$}|}{f=5} \\   
  \hline n &j=2 &j=3 &j=4 &j=5  \\
  \hline 2 &91 &155 &235 &331   \\
  \hline 3 &455 &739 &1,071 &1,451   \\
  \hline 4 &3,647 &6,601 &10,419 &15,101   \\
  \hline 5 &23,255 &40,157 &61,059 &85,961   \\ 
  \hline 6 &200,473 &372,719 &598,005 &876,331   \\
  \hline 7 &1,470,061 &2,636,375 &4,118,865 &5,917,531   \\
  \hline 8 &13,229,719 &24,937,645 &40,333,059 &59,415,961   
\\\hline
\end{tabular}
\end{center}

\subsection{Gluonic contributions}
An interesting question that arises is the issue of contribution of gluonic processes, 
compared to the total number of processes, since often the purely gluonic process
is assumed to be typical or 'dominant'. In particular we would like to examine to 
what degree purely gluonic amplitudes dominate over other kinds of processes, since
gluons have a different color charge than quarks \footnote{Note that we do 
\underline{not} address the question of the singularity structure of gluonic versus
other amplitudes.}. To this 
end we assign to each external gluon an additional factor $k$, resulting in a 
modification of the generating function (\ref{Gen-Fun}):
\bq
\mathcal{G}_k(x)=e^{kx} I_0(2x)^{f-j} \left( k+\frac{d}{dx} \right)^2 I_0(2x)^j
\eq
The corresponding generating function for purely gluonic processes would be:
\bq
\mathcal{G}_k^0(x)=k^2 e^{kx}
\eq
We may compare the coefficients of the expansion of the two generating functions.
This can be seen in the graph that follows, where we have plotted the ratio
$ \mathcal{G}_{k,n}^0 / \mathcal{G}_{k,n}$ of the coefficients, for number of
jets ranging from $n=2$ up to $n=8$, against $k$, and for the particular case of
$f=3,j=3$. Notice that the ratio approaches one as the factor $k$ grows larger,
but decreases with $n$.
\begin{center}
\epsfig{file=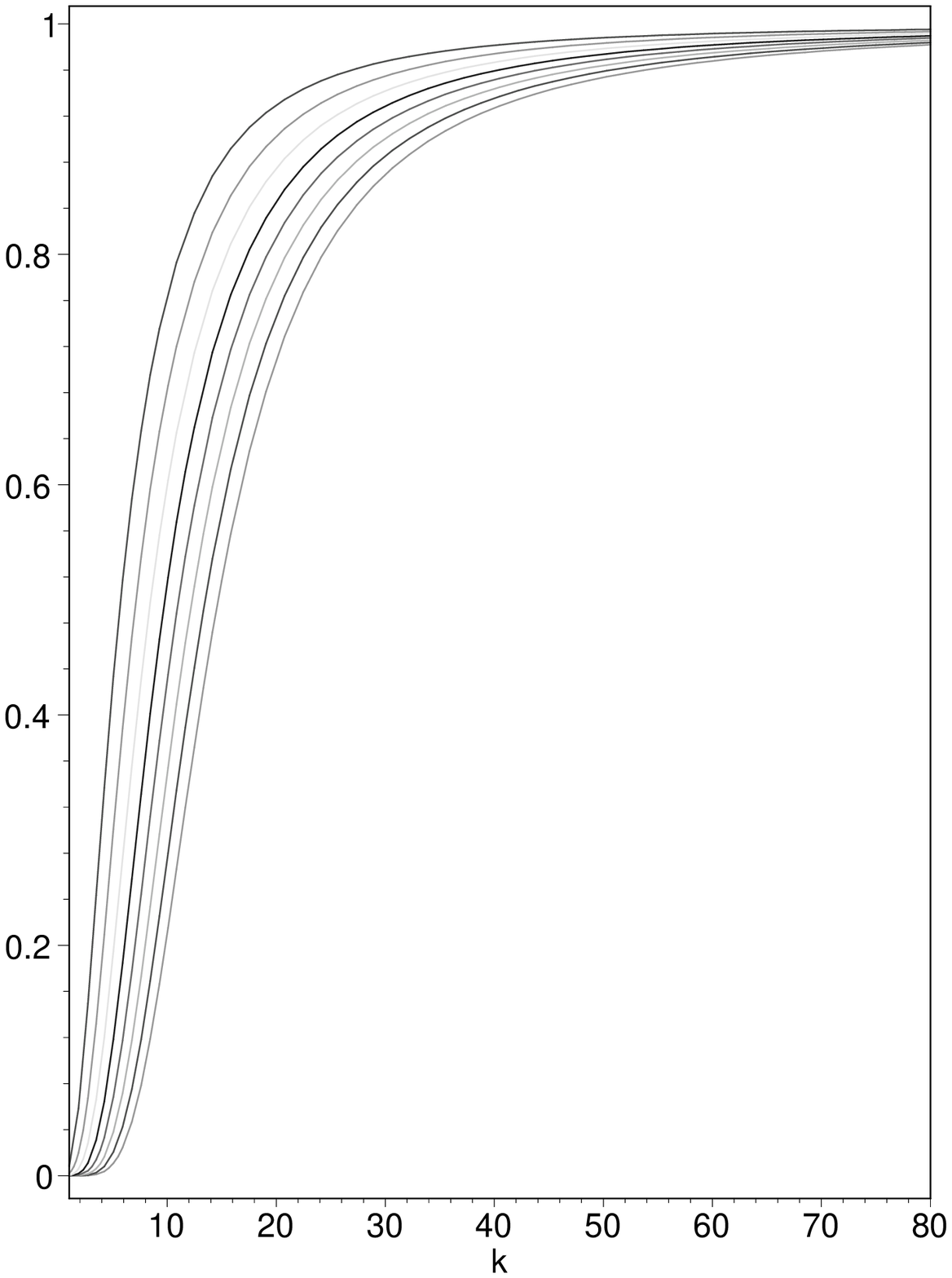,height=6.8cm,width=6.8cm}
\end{center}
In order to estimate how large $k$ has to make the gluonic amplitude the dominant one,
we look for values of $k$ that give $ \mathcal{G}_{k,n}^0 / \mathcal{G}_{k,n}=1/2$. 
These can be seen in the table
that follows, for the case $f=3,j=3$ again and for various numbers of jets.
\begin{center}
\begin{tabular}{|c||c|c|c|c|c|c|c|c|}
\hline
n & 2 & 3 & 4 & 5 & 6 & 7 & 8 & 9  \\
\hline
$ k $ & 5.72 & 7.13 & 8.47 & 9.78 & 11.04 & 12.27 & 13.48 & 14.67 \\
\hline
\end{tabular} 
\end{center}
Another extension would be to account for the fact that the gluonic structure function
is typically larger than that for a quark. Then the
generating function becomes:
\bq
\mathcal{G}_S(x)=e^{x} I_0(2x)^{f-j} \left( S+\frac{d}{dx} \right)^2 I_0(2x)^j
\eq
where $S$ denotes the gluon structure function enhancement factor. We see that this is a 
monotonically increasing, quadratic function of $S$. For large $S$ the generating function
can be approximated by
\bq
\mathcal{G}_S(x)=S^2 e^{x} I_0(2x)^{f} 
\eq
The coefficients of $\mathcal{G}_S(x)$ may be written $\mathcal{G}_{n}=
C_n S^2$ where the $C_n$  depend on $n$. We can estimate the
'strength' of the 'S-extended' gluonic amplitudes compared to purely gluonic processes, 
$ \mathcal{G}_{n}^0=S^2$, by computing the ratio $r=\lim_{s \to \infty} 
\mathcal{G}_{n}^0 / \mathcal{G}_{n}$.
These ratios are tabulated below:
\begin{center}
\begin{tabular}{|>{$}c<{$}||>{$}c<{$}|>{$}c<{$}|>{$}c<{$}|}
  \hline \multicolumn{4}{|>{$}c<{$}|}{\mathrm{The \; \; ratio \; \; \mathcal{G}_{n}^0 / \mathcal{G}_{n} }}\\
  \hline n &f=3 &f=4 &f=5   \\
  \hline 2 &0.1428 &0.1111 &0.0909    \\
  \hline 3 &0.0526 &0.40 &0.0322    \\
  \hline 4 &0.0078 &0.0046 &0.0030    \\
  \hline 5 &0.0019 &0.00108 &0.00068    \\ 
  \hline 6 &0.0003 &0.00012 &0.000066   \\
  \hline 7 &0.000061 &0.000023 &0.000011    \\
  \hline 8 &0.968 10^{-5} &0.289 10^{-5} &0.114 10^{-5}   
\\\hline
\end{tabular}
\end{center}
We conclude that for sizeable $n$ the purely gluonic amplitudegives only a 
very small contribution.

\subsection{Asymptotic results}
It may be interesting to estimate the number of amplitudes for large number of
jets. To this end, we would like to obtain the asymptotic form of the generating
function for large $n$. The asymptotic expansion for $I_0(2z)$ is
\bq
I_0(2z) \sim {e^{2z}\over\sqrt{4\pi z}}\suml_{n\ge0}{\tau_n\over z^n}\;\;\;,
\;\;\;\tau_n = {(2n)!^2\over 64^n\;n!^3}\;\;\;,\;\;\; z \to \infty.
\eq
This expansion holds for Re$(z)>0$, but we also have $I_0(-z)=I_0(z)$.
For the function
\bq
f(x)=I_0(2x)^p
\eq
 the asymptotic expansion is
\bq
f(x)=N e^{2px} x^{- \frac{p}{2} } \sum_{n \geq 0} \frac{ \alpha_n }{x^n}
\; \; \;, \; \; \alpha_n=\sum_{n_1,\ldots,n_p} \tau_{n_1} \tau_{n_2} \cdots 
\tau_{n_p} \Theta(n_1+\cdots+n_p=n)
\eq
where $N=(4 \pi)^{- \frac{p}{2} }$. 
So the derivatives in the generating function $\mathcal{G}(x)$, read:
\bq
\left( 1+\frac{d}{dx} \right)^2 f(x)=f(x)+2 f^{\prime}(x) + f^{\prime \prime}(x)=
N e^{2px} x^{- \frac{p}{2} } \sum_{n \geq 0} \frac{ \beta_n }{x^n}
\eq
where
\bq
\beta_n=(1+2p)^2 \alpha_n-2(1+2p)(n+\frac{p}{2}-1) \alpha_{n-1} 
+ (n+\frac{p}{2}-1)(n+\frac{p}{2}-2) \alpha_{n-2},
\eq
and for the generating function we get
\bq
\mathcal{G}(x)=\frac{1}{(4 \pi)^{f/2}} \; e^{x(1+2f)} \; \frac{1}{x^{f/2}} \;
\sum_{n \geq 0} \frac{\gamma_n}{x^n}
\label{asymexpr}
\eq
where
\bq
\gamma_n=\sum_{n_1,n_2 \geq 0} \Theta(n_1+n_2=n) \; \alpha_{n_1}(f-j) \; \beta_{n_2}(j)
\eq
The first few $\gamma$'s for various numbers of initial and final flavours, are
shown in the next table. 

\begin{center}
\begin{tabular}{|>{$}c<{$}|>{$}c<{$}|>{$}c<{$}|>{$}c<{$}|>{$}c<{$}|>{$}c<{$}|}
  \hline \multicolumn{2}{|>{$}c<{$}|}{\mathrm{Number \; of \; flavours} } &
         \multicolumn{1}{>{$}c<{$}|}{\gamma_0 } & 
         \multicolumn{1}{>{$}c<{$}|}{\gamma_1 } & 
         \multicolumn{1}{>{$}c<{$}|}{\gamma_2 } & 
         \multicolumn{1}{>{$}c<{$}|}{\gamma_3 } \\ 
  \hline \multirow{4}{10mm}[4mm]{$f=3$}
            & \multirow{2}{10mm}[2.6mm]{$j=2$}
               & 25 & -85/16 & 249/512 & 1873/8192 \\\cline{2-6}
            & \multirow{2}{10mm}[2.6mm]{$j=3$}
               & 49 & -189/16 & 177/512 & 1337/8192 \\\hline
         \multirow{3}{10mm}{$f=4$}
            & \multirow{2}{10mm}[2.6mm]{$j=2$}
               & 25 & -15/4 & 19/32 & 101/256 \\\cline{2-6}
            & \multirow{2}{10mm}[2.6mm]{$j=3$}
               & 49 & -35/4 & 15/32 & 109/256 \\\cline{2-6}
            & \multirow{2}{10mm}[2.6mm]{$j=4$}
               & 81 & -63/4 & 3/32 & 93/256 \\\hline
         \multirow{4}{10mm}{$f=5$}   
            & \multirow{2}{10mm}[2.6mm]{$j=2$}
               & 25 & -35/16 & 409/512 & 4871/8192 \\\cline{2-6}
            & \multirow{2}{10mm}[2.6mm]{$j=3$}
               & 49 & -91/16 & 401/512 & 6143/8192 \\\cline{2-6}
            & \multirow{2}{10mm}[2.6mm]{$j=4$}
               & 81 & -171/16 & 273/512 & 6831/8192 \\\cline{2-6}
            & \multirow{2}{10mm}[2.6mm]{$j=5$}
               & 121 & -275/16 & 25/512 & 6935/8192 \\\hline
\end{tabular}
\end{center}
The most important term in the series is of course the first 
\bq
\gamma_0=\alpha_{0}(f-j) \; \beta_{0}(j)=\alpha_{0}(f-j) \; (1+2j)^2 \; \alpha_{0}(j)
        =(1+2j)^2
\eq
Keeping only this term in the generating function we have
\bq
\mathcal{G}(x) \sim \frac{1}{(4 \pi)^{f/2}} \; e^{x(1+2f)} \; \frac{1}{x^{f/2}} \; (1+2j)^2
\; \; , \; \; x \to \infty
\eq

Let us, now, assume that we want to include the first $K$ terms in the
asymptotic expansion of $\mathcal{G}(x)$, that is, we set $\gamma_j$ to zero for
$j>K$. The Borel transform
\bq
\mathcal{F}(x) = \intl_0^\infty\;dy\;y^{K+f/2}e^{-y} \mathcal{G}(xy)
\label{intrep}
\eq
has the expansion
\bq
\mathcal{F}(x) = \suml_{n\ge0} \Gamma(n+K+f/2+1)\mathcal{G}_n\;x^n\;\;,
\eq
where $\mathcal{G}(x)=\sum_{n \geq 0} \mathcal{G}_n x^n$; 
our approach consists in finding the coefficients of $\mathcal{F}(x)$ by studying
its singularity structure. The integral for $\mathcal{F}(x)$ can be written as
\bqa
\mathcal{F}(x) &=& {1\over(4\pi)^{f/2}}
\intl_0^\infty\;dy\;\exp[(-y+xy(2f+1))]\suml_{k=0}^K
\gamma_k{y^{K-k}\over x^{k+f/2}}\nl
&=& {1\over(4\pi)^{f/2}}
\suml_{k=0}^K{\gamma_k(K-k)!\over x^{k+f/2}(1-x(2f+1))^{K-k+1}}\;\;.
\eqa
This expression is has a pole at $x_0=1/(2f+1)$. Note that, due to our
use of the factor $y^{K+f/2}$, the integral (\ref{intrep}) is indeed
dominated by large values of $xy$ when $x$ approaches $x_0$, thus justifying
the use of the asymptotic expression (\ref{asymexpr}). Furthermore, 
there is of course a similar singularity which appears when we use
negative $x$ values: however, since that is located at $-1/(2f-1)$ and
hence further away from the origin than $x_0$, this pole will give
exponentially suppressed contributions which will not show up
in our result for $\mathcal{G}_n$. The $k^{\mbox{\small th}}$
term in the series for $\mathcal{F}(x)$ is seen to contain poles at $x=x_0$
of order up to and including $K-k+1$:
\bqa
\lefteqn{ x^{-k-f/2}(1-{x\over x_0})^{-K+k-1} =}\nl
&& {1\over x_0^{k+f/2}}\suml_{r=0}^{K-k}
{(k+f/2+r)!\over r!(k+f/2-1)!}(1-{x\over x_0})^{-K+k+r-1}\;\;
+\;\;\mbox{regular terms}\;\;.
\eqa
The dominant behaviour of the coefficient of $x^n$ in the series
expansion of this term is, therefore,
\bqa
\lefteqn{ {1\over x_0^{n+k+f/2}}\suml_{r=0}^{K-k}\;
{(k+f/2+r)!\over r!(k+f/2-1)!}
{(n+K-k-r)!\over n!(K-k-r)!} =}\nl
&& {1\over x_0^{n+k+f/2}}{(n+K+f/2)!\over(K-k)!(n+k+f/2)!}\;\;.
\eqa
Inserting this in the expression for $\mathcal{F}(x)$ and dividing by the factor
$\Gamma(n+K+f/2+1)$ to get the coefficient $\mathcal{G}_n$, we see that $K$
drops out from the expression, so that we may take it as large as we please.
The resultant form for $\mathcal{G}_n$ is, therefore
\bq
\mathcal{G}_n \sim \mathcal{G}_n^{asy} ={(2f+1)^{n+f/2}\over(4\pi)^{f/2}}
\suml_{k\ge0} {\gamma_k(2f+1)^k\over\Gamma(n+k+f/2+1)}\;\; , \; \; n \to \infty.
\eq
In order to estimate how accurate this asymptotic expansion is, we have
calculated the ratio between the exact and the "asymptotic" number of processes.
The results are shown in the next table, where we have recorded the way these
numbers improve as we add more terms in the asymptotic expansion of the generating
function. Thus $n_0$ is such that $ \mathcal{G}_n/\mathcal{G}_n^{asy}$ 
is between $0.95$ and $1.05$ for all $n \geq n_0$, when we include only the first
term in the expansion, i.e. the term that contains $\gamma_0$. Similarily, $n_1$
is the number of jets when we include $\gamma_0$ and $\gamma_1$, $n_2$ when
we include $\gamma_0$, $\gamma_1$ and $\gamma_2$, etc.

\begin{center}
\begin{tabular}{|>{$}c<{$}|>{$}c<{$}|>{$}c<{$}|>{$}c<{$}|>{$}c<{$}|>{$}c<{$}|}
  \hline \multicolumn{2}{|>{$}c<{$}|}{\mathrm{Number \; of \; flavours} } &
         \multicolumn{1}{>{$}c<{$}|}{n_0 } & 
         \multicolumn{1}{>{$}c<{$}|}{n_1 } & 
         \multicolumn{1}{>{$}c<{$}|}{n_2 } & 
         \multicolumn{1}{>{$}c<{$}|}{n_3 } \\ 
  \hline \multirow{4}{10mm}[4mm]{$f=3$}
            & \multirow{2}{10mm}[2.6mm]{$j=2$}
               & 26 & 6 & 6 & 6 \\\cline{2-6}
            & \multirow{2}{10mm}[2.6mm]{$j=3$}
               & 31 & 5 & 4 & 5 \\\hline
         \multirow{3}{10mm}{$f=4$}
            & \multirow{2}{10mm}[2.6mm]{$j=2$}
               & 21 & 8 & 8 & 6 \\\cline{2-6}
            & \multirow{2}{10mm}[2.6mm]{$j=3$}
               & 28 & 8 & 6 & 6 \\\cline{2-6}
            & \multirow{2}{10mm}[2.6mm]{$j=4$}
               & 31 & 7 & 7 & 7 \\\hline
         \multirow{4}{10mm}{$f=5$}   
            & \multirow{2}{10mm}[2.6mm]{$j=2$}
               & 11 & 12 & 10 & 8 \\\cline{2-6}
            & \multirow{2}{10mm}[2.6mm]{$j=3$}
               & 21 & 12 & 10 & 10 \\\cline{2-6}
            & \multirow{2}{10mm}[2.6mm]{$j=4$}
               & 25 & 12 & 10 & 10 \\\cline{2-6}
            & \multirow{2}{10mm}[2.6mm]{$j=5$}
               & 29 & 12 & 11 & 10 \\\hline
\end{tabular}
\end{center}
Note that on some occasions, like for example $f=5, \; j=2$, the number  
increases as we add more terms in the expansion. But this is due to a small increase
of the ratio which is improved immediately when we add the next term.

\section{Computational complexity: distinct amplitudes}
In the above we have shown how {\em all\/} amplitudes contributing
to a certain cross section can be enumerated. This would also, then, be
the computational complexity in an approach where each amplitude is
calculated from scratch. However, there is of course a simplification
owing to the fact that amplitudes that differ only by a relabelling of the
(massless!) quark flavours are equal apart from a trivial difference
in the structure function. It therefore behooves us to take this
simplification into account. Now it must be kept in mind that, when a quark
flavour occurs in the initial state, we should not relabel it since that is
also taken care of by the factors $2j$, $j(j-1)$ etcetera in table 1.
Only those quark flavours that do not occur in the initial state may be
relabelled. Let us perform the relabelling in such a way that the
relabelled flavours occur in order of increasing multiplicity. As
an example, in the process $gg\to X$ this means that instead of
\bq
A(x) = \suml_{n_{0,1,\ldots,f}\ge0} {n!x^{n_0+2(n_1+\cdots+n_f)}\over
n_0!(n_1!)^2\cdots(n_f!)^2}\;\;,
\eq
we have to determine, rather,
\bq
\tilde{A}_f(x) = \suml_{n_0\ge0}\suml_{0\le n_1\le n_2\le\cdots\le n_f}
 {n!x^{n_0+2(n_1+\cdots+n_f)}\over n_0!(n_1!)^2\cdots(n_f!)^2}\;\;;
\eq
likewise, for the process $gq\to X$ we have to compute
\bq
\tilde{B}_f(x) = 
\suml_{n_{0,1}\ge0}\suml_{0\le n_2\le n_3\le\cdots\le n_f}
 {n!x^{n_0+1+2(n_1+\cdots+n_f)}\over n_0!
n_1!(n_1+1)!(n_2!)^2\cdots(n_f!)^2}\;\;.
\eq
It is important to note that those quark flavours that can be relabelled occur
symmetrically in these sums.

The straightforward implementation of such inequalities appears to lead to
horrendous complications. An exception is the following
generating function:
\bq
Z_f(x) = \suml_{0\le n_1\le n_2\le\cdots\le n_f}x^{n_1+n_2+\cdots+n_f}
= {1\over(1-x)(1-x^2)(1-x^3)\cdots(1-x^f)}\;\;,
\eq
familiar from the theory of partitions \cite{Hardy}.
In fact, we may employ the symmetry in the relabelled 
indices. To see how this works, let us symmetrize the case $f=2$:
\bq
\theta(n_1\le n_2) \rightarrow
{1\over2}(\theta(n_1\le n_2) + \theta(n_2\le n_1))
= {1\over2}(1+\theta(n_1=n_2))\;\;.
\eq
This can be obviously extended to larger $f$: the inequalities lead
to a combination of terms with no restriction, terms where two labels are
equated, terms where three labels are equated, terms where four labels are
grouped in two pairs of equal ones, and so on. Using the function $Z(x)$, 
we can convieniently determine the various coefficients by working out
how $Z_f(x)$ can be split up in the corresponding way. Again for the case
$f=2$, this means writing
\bq
Z_2(x) = {\alpha\over(1-x)^2} + {\beta\over(1-x^2)}\;\;,
\eq
and solving this for general $x$ gives, indeed, $\alpha=\beta=1/2$.
As usual in the theory of partitions, a result for general $f$ is 
prohibitively complicated, and therefore we give only the first 
few values of $f$:
\bqa
Z_2(x) &=& {1\over2}{1\over(1-x)^2} + {1\over2}{1\over(1-x^2)}\;\;,\nl
Z_3(x) &=& {1\over6}{1\over(1-x)^3} + {1\over2}{1\over(1-x)(1-x^2)}
 + {1\over3}{1\over(1-x^3)}\;\;,\nl
Z_4(x) &=& {1\over24}{1\over(1-x)^4} + {1\over4}{1\over(1-x)^2(1-x^2)}
 + {1\over3}{1\over(1-x)(1-x^3)}\nl
&& + {1\over8}{1\over(1-x^2)^2} + {1\over4}{1\over(1-x^4)}\;\;,\nl
Z_5(x) &=& {1\over120}{1\over(1-x)^5} + {1\over12}{1\over(1-x)^3(1-x^2)}
 + {1\over6}{1\over(1-x)^2(1-x^3)}\nl
&& + {1\over4}{1\over(1-x)(1-x^4)} + {1\over8}{1\over(1-x)(1-x^2)^2}
+ {1\over6}{1\over(1-x^2)(1-x^3)}\nl && + {1\over5}{1\over(1-x^5)}\;\;.
\eqa
The result for $\tilde{A}(x)$ in these cases is therefore:
\bqa
\tilde{A}_1(x) &=&
e^xH_2(x)\;\;,\nl
\tilde{A}_2(x) &=&
{e^x\over2}\left( H_2(x)^2 + H_4(x) \right)\;\;,\nl
\tilde{A}_3(x) &=&
{e^x\over6}\left( H_2(x)^3 + 3H_2(x)H_4(x) + 2H_6(x) \right)\;\;,\nl
\tilde{A}_4(x) &=&
{e^x\over24}\left( H_2(x)^4 + 6H_2(x)^2H_4(x) + 8H_2(x)H_6(x) \right.\nl
&&\hphantom{{e^x\over24}}\left.
+ 3H_4(x)^2 + 6H_8(x)\right)\;\;,\nl
\tilde{A}_5(x) &=&
{e^x\over120}\left( H_2(x)^5 + 10H_2(x)^3H_4(x) + 20H_2(x)^2H_6(x)
+ 30H_2(x)H_8(x)\right.\nl
 &&\hphantom{{e^x\over240}}\left.
 + 15H_2(x)H_4(x)^2 + 20H_4(x)H_6(x) + 24H_{10}(x)\right)\;\;.
 \label{Af}
\eqa
These identities can easily be checked explicitly to modest order in $x$.
Here, we have introduced the class of generalized hypergeometric functions
\bq
H_m(x) = \suml_{n\ge0}\left( {x^n\over n!}\right)^m\;\;.
\eq
Obviously, $H_1(x)=e^x$, while $H_2(x)=I_0(2x)$.
The rest of the cases are treated in a similar fashion. 
The number of dinstinct amplitudes that need to be calculated contains:
\bqa
\tilde{\mathcal{B}}(x)&=&I^{\prime}_0(2x) \tilde{A}_{f-1}(x) \label{B}\\
\tilde{\mathcal{C}}(x)&=&\tilde{A}_{f-1}(x)  \cdot 
\{ 2 I_0^{\prime \prime}(2x)-I_0(2x) \} \\
\tilde{\mathcal{D}}(x)&=&\tilde{A}_{f-2}(x) \left( I^{\prime}_0(2x) \right)^2
\label{D}
\eqa
The generating function $\tilde{\mathcal{G}}$ for the dinstict amplitudes can be written
as in (\ref{Gen-Fun-alt}):
\bq
\tilde{\mathcal{G}}(x)=\tilde{\mathcal{A}}_f(x)+2I_0(2x)\tilde{\mathcal{A}}_{f-1}(x)
+4 \tilde{\mathcal{B}}(x) +2 \tilde{\mathcal{C}}(x) +6 \tilde{\mathcal{D}}(x) 
\eq 
Note the occurence of coefficients 4 and 6, which are dictated by a careful examination
of the differing initial states that contribute.
Some numbers for the case of $f=3,4,5$ flavours, are shown  in the following table

\begin{center}
\begin{tabular}{|>{$}c<{$}||>{$}c<{$}|>{$}c<{$}||>{$}c<{$}|}
  \hline \multicolumn{4}{|>{$}c<{$}|}{Total \; number \; of \; dinstinct \; amplitudes }\\
  \hline n &f=3 &f=4 &f=5  \\
  \hline 2 &35 &35 &35   \\
  \hline 3 &123 &123 &123    \\
  \hline 4 &777 &777 &777    \\
  \hline 5 &3,853 &3,853 &3,853    \\ 
  \hline 6 &25,327 &31,087 &31,087   \\
  \hline 7 &139,975 &200,455 &200,455   \\
  \hline 8 &870,485 &1,676,885 &1,999,445    
\\\hline
\end{tabular}
\end{center}
Note that for small $n$ the numbers coincide: this is due to the fact that, $n=4$, say,
allows no room for 4 different quark flavours to occur in one diagram, and the difference 
between $f=3$ and $f=4$ can therefore only appear for $n \geq 6$. This is reflected 
in the fact that $\tilde{\mathcal{A}}_f(x)$ coincides with  $\tilde{\mathcal{A}}_{f-1}(x)$ up
to the $x^{2f}$ term. 
\newline

In order to estimate the large $x$ expansion of the generating function, as 
in section 2, we note that $H_f(x)$ is
\bq
H_f(x) \sim \frac{e^{2xf}}{(4 \pi )^{1/2} x^{1/2}} \left( 1+\mathcal{O}(\frac{1}{x}) \right)
\; \; , \; \; x \to \infty
\eq
In the Appendix we show how the asymptotic expansion can be computed systematically
and using this, we can approximate $\tilde{\mathcal{A}}_f(x)$ by
\bq
\tilde{\mathcal{A}}_f(x)=\frac{1}{f!}H_2^f(x)+\frac{1}{2(f-2)!}H_2^{f-2}(x)H_4(x)
\eq
taking into account equations (\ref{Af}) and keeping only the largest powers of $x$.
The second term in the previous equation, gives $\frac{1}{\sqrt{x}}$ corrections
to the leading result. Using this and the derivatives of the Bessel function
\bq
I_0^{ \prime }(2x)\;\; , \;\;I_0^{ \prime \prime }(2x) \sim
 H_2(x) \left(1+\mathcal{O}(\frac{1}{x}) \right) \; \; , \; \; x \to \infty
\eq
we can estimate the large $x$ expansion of the functions in (\ref{B}-\ref{D}):
\bqa
\tilde{\mathcal{B}}(x)& \sim &\tilde{\mathcal{A}}_{f-1}(x) H_2(x)=\frac{1}{(f-1)!}H_2^f(x)
+\frac{1}{2(f-3)!}H_2^{f-2}(x)H_4(x) \\
\tilde{\mathcal{C}}(x)& \sim &\tilde{\mathcal{A}}_{f-1}(x) \left( I_0^{ \prime \prime }(2x)
-I_0(2x) \right)=\tilde{\mathcal{A}}_{f-1}(x) H_2(x)=\tilde{\mathcal{B}}(x) \\
\tilde{\mathcal{D}}(x)& \sim &\tilde{\mathcal{A}}_{f-2}(x)(I_0^{ ' })^2=
\frac{1}{(f-2)!}H_2^f(x)+\frac{1}{2(f-4)!}H_2^{f-2}(x)H_4(x)
\eqa
and the generating function:
\bq
\tilde{\mathcal{G}}(x) \sim \frac{1+2f+6f^2}{f!}H_2^f(x)+ \frac{6f^2-22f+2}{2(f-2)!} H_2^{f-2}(x)
H_4(x)
\label{Gen-din}
\eq
We can also compute the coefficients of the asymptotic expansion of $\tilde{\mathcal{G}}(x)$.
To this end we calculate the coefficient in the expansion of the functions $H_2(x) , H_4(x)$
using the Borel transform. In particular for $H_2(x)$ we define
\bq
P(x)=e^xH_2^f(x)=\sum_n K_n x^n
\eq
To estimate the coefficients $K_n$ we perform a transform on $P(x)$:
\bq
\int_0^{\infty} dy e^{-y}y^{f/2} P(xy)= \sum_n K_n \Gamma(n+\frac{f}{2}+1) x^n 
\eq
and we get
\bq
K_n \sim \frac{1}{(4 \pi )^{f/2}} \frac{(1+2f)^{n+f/2}}{\Gamma(n+\frac{f}{2}+1)}
\eq
Similarily, for $H_4(x)$, approximated by
\bq
H_4(x) \sim \frac{e^{4x}}{(32 \pi^3)^{1/2} x^{3/2}}\; \; , \; \; x \to \infty
\eq
we use $Q(x)=e^x H_2^{f-2}(x)H_4(x)=\sum_n L_n x^n$. Performing a Borel transform we get
\bq
\int_0^{\infty} dy e^{-y}y^{\frac{f}{2}+\frac{1}{2}} Q(xy)= 
\sum_n L_n \Gamma(n+\frac{f}{2}+\frac{3}{2}) x^n
\eq
and for the coefficients
\bq
L_n \sim \frac{1}{(4 \pi )^{f/2}} 
\frac{(1+2f)^{n+f/2+1/2}}{\sqrt{2 \pi} \Gamma(n+\frac{f}{2}+\frac{3}{2})}
=K_n \left( \frac{1+2f}{2 \pi} \right)^{1/2} \frac{\Gamma(n+\frac{f}{2}+1)}
{\Gamma(n+\frac{f}{2}+\frac{3}{2})}
\eq
Using these coefficients we can estimate the coefficients for the generating function
in (\ref{Gen-din}).

\vspace*{1cm}

\section*{{\LARGE Appendix: Asymptotic form of $H_m(x)$}}




Here we study the asymptotic form of the function $H_m(x)$,
 which was defined as
\bq
H_m(x)=\sum_{n \geq 0} \frac{x^{mn}}{(n!)^m}
\eq
One can easily see that the following relation holds
\bqa
H_m(x)&=& \frac{1}{2 \pi i} \oint \frac{dz}{z}H_{m-1}(xz)H_1(\frac{x}{z}) \nn \\
      &=& \frac{1}{(2 \pi i)^{m-1}} \oint \cdots \oint \frac{dz_1}{z_1} \cdots 
      \frac{dz_{m-1}}{z_{m-1}} H_1(xz_1)H_1(xz_2) \cdots H_1(\frac{x}{z_1z_2 \cdots z_{m-1}})
\eqa
If we put $z_i=e^{i \phi_i}$ the integral becomes
\bq
H_m(x)=\frac{1}{(2 \pi)^{m-1}} \int_0^{2\pi} \cdots \int_0^{2\pi}
d \phi_1 \cdots d \phi_{m-1} e^{xW}
\eq
where
\bq
W=e^{i \phi_1}+ \cdots e^{i \phi_{m-1}}+e^{-i ( \phi_1+ \cdots \phi_{m-1} )}
\eq
We can estimate this integral by using the saddle point approximation. The first few 
derivatives of $W$ are
\bqa
\frac{ \partial W}{\partial \phi_k}&=& i \left( e^{i \phi_k}-e^{-i(\phi_1+ \cdots \phi_{m-1} )} \right)\; , \nn\\
\frac{ \partial^2 W}{\partial \phi_k \partial \phi_{\ell} }&=&
 -\left( e^{i \phi_k} \delta_{k \ell}+e^{-i(\phi_1+ \cdots \phi_{m-1} )} \right)\; , \nn \\
\frac{ \partial^3 W}{\partial \phi_k \partial \phi_{\ell} \partial \phi_p}&=&
 - i \left( e^{i \phi_k} \delta_{k \ell p}-e^{-i(\phi_1+ \cdots \phi_{m-1} )} \right) \; , \nn \\
\frac{ \partial^4 W}{\partial \phi_k \partial \phi_{\ell} \partial \phi_p \partial \phi_q}&=&
 \left( e^{i \phi_k} \delta_{k \ell p q}+e^{-i(\phi_1+ \cdots \phi_{m-1} )} \right)
 \; \; , \ldots 
\eqa 
The saddle point can be found from the first derivative, and it is the solution of the
equation
\bq
e^{i \phi}+e^{-i(m-1)\phi}=0 \to e^{im\phi}=1 \to \phi=\frac{2 \pi}{m} k, \; \; k=0,1,\ldots m-1
\eq
The value of $xW$ at the  saddle point is $x((m-1)e^{i\phi}+e^{i\phi})=mxe^{i\phi}$. 
The saddle point that gives the largest real part of $mxe^{i\phi}$ is the one that 
dominates. We see that the function has an $m$-fold symmetry: if we restrict ourselves to 
$|arg(x)| < \frac{\pi}{m}$ the saddle point that dominates is $\phi=0$. 
The derivatives now take the values:
\bqa
\frac{ \partial^2 W}{\partial \phi_k \partial \phi_{\ell} }&=& -\left( \delta_{k \ell} +1 \right) \; , \nn \\
\frac{ \partial^3 W}{\partial \phi_k \partial \phi_{\ell} \partial \phi_p}&=&
-i \left( \delta_{k \ell p} -1 \right)  \; , \nn \\
\frac{ \partial^4 W}{\partial \phi_k \partial \phi_{\ell} \partial \phi_p \partial \phi_q}&=&
\left( \delta_{k \ell p q} +1 \right) \; \; , \ldots 
\eqa
and the exponent is
\bq
xW=mx-\frac{x}{2} \sum_{k \ell} \left( \delta_{k \ell} +1 \right) \phi_k \phi_{\ell}
-\frac{ix}{6} \sum_{k \ell p} \left( \delta_{k \ell p} -1 \right) \phi_k \phi_{\ell} \phi_p
+\frac{x}{24} \sum_{k \ell p q} \left( \delta_{k \ell p q} +1 \right) \phi_k \phi_{\ell} \phi_p \phi_q
+ \cdots
\eq
This is reminiscent of a zero-dimensional scalar field theory with vertices of arbitrary
multiplicity, with the Feynman rules
\bqa
&& \frac{1}{x} \left( \delta_{\mu \nu}-\frac{1}{m} \right) 
\; \; \; \; \; \; \;  \epsfig{file=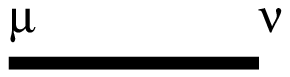,width=1.5cm} \; \; \; \; ,  \nn \\
&& -ix \left( \delta_{\mu \nu \alpha}-1 \right) 
\; \; \; \; \; \; \;  \; \; \raisebox{-10mm}{\epsfig{file=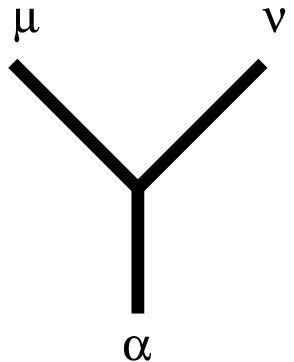,width=1.5cm}}\; \; \; \;,  \nn \\
&& x \left( \delta_{\mu \nu \alpha \beta}+1 \right) 
\; \; \; \; \; \; \;   \raisebox{-10mm}{\epsfig{file=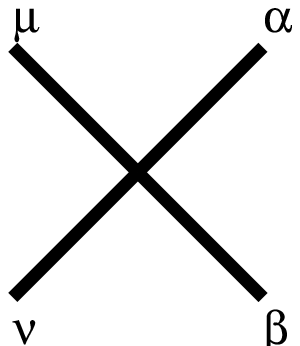,width=1.5cm}}\; \;\; \; ,  \nn \\
&& ix \left( \delta_{\mu \nu \alpha \beta \rho}-1 \right)
\; \; \; \; \; \; \;   \raisebox{-10mm}{\epsfig{file=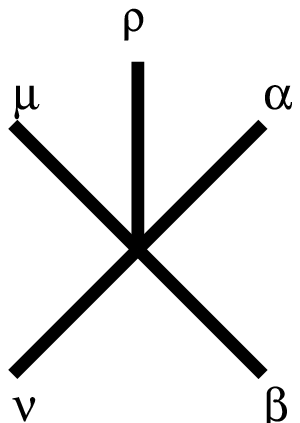,width=1.5cm}}\; \; \; \;,  \nn \\
&& -x \left( \delta_{\mu \nu \alpha \beta \rho \sigma}+1 \right)
\; \; \; \; \; \; \;   \raisebox{-10mm}{\epsfig{file=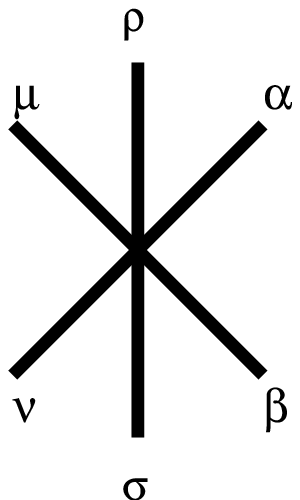,width=1.5cm}}\; \; \; \;, \ldots \nn 
\eqa
where $\delta_{\mu \nu \alpha}=\delta_{\mu \nu}\delta_{\mu \alpha}, \; 
\delta_{\mu \nu \alpha \beta}=\delta_{\mu \nu} \delta_{\mu \alpha} \delta_{\mu \beta}$,
and so on.
We can use the familiar tools of field theory to evaluate the integral. The first
subleading term is computed by taking into account the two-loop diagrams that contribute.
The result is
\bq
\frac{1}{8} \; \raisebox{-2mm}{\epsfig{file=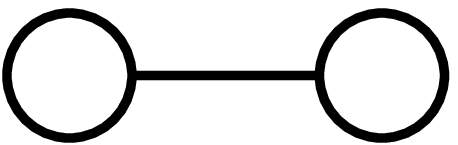,width=2cm}}
+\frac{1}{8} \; \raisebox{-2mm}{\epsfig{file=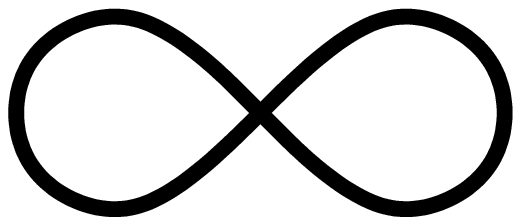,width=1.3cm}} 
+ \frac{1}{12} \; \raisebox{-4mm}{\epsfig{file=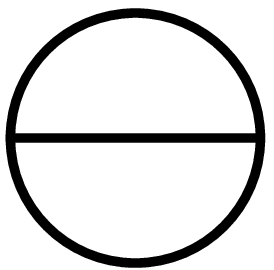,width=1cm}}
=0+\frac{(m-1)^2}{8mx}-\frac{(m-1)(m-2)}{12mx}=\frac{m^2-1}{24mx}
\eq
where the factors in front of the diagrams are symmetry factors. The next subleading term 
can be computed by including three-loop graphs. Due to the fact that
$\raisebox{-1.8mm}{\epsfig{file=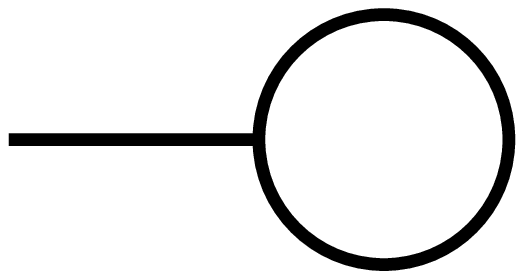,width=1cm}}=0$, there are 
8 non-zero connected three-loop diagrams, and in addition to these
we must also include the disconnected diagrams that are shown below. The 
result for the next term in the expansion of $H_m(x)$ (including the symmetry factors
shown below) is:
\bqa
&+&\frac{1}{48} \; \; \raisebox{-4mm}{\epsfig{file=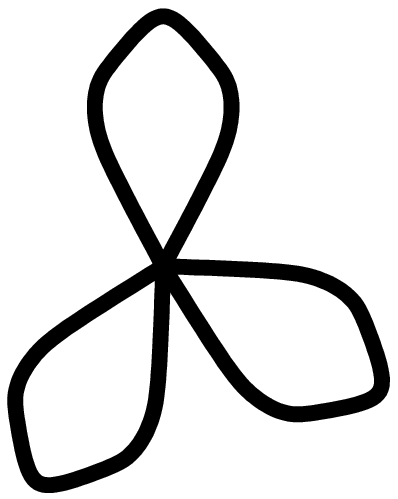,width=0.8cm}}
+\frac{1}{12} \; \; \raisebox{-2mm}{\epsfig{file=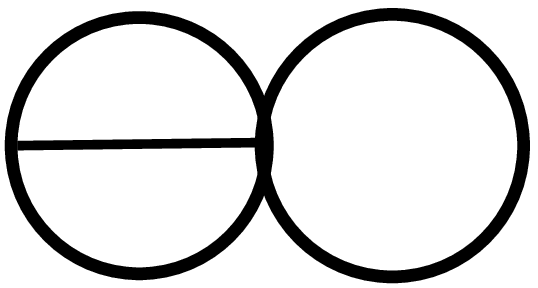,width=1.2cm}}
+\frac{1}{48} \; \; \raisebox{-3mm}{\epsfig{file=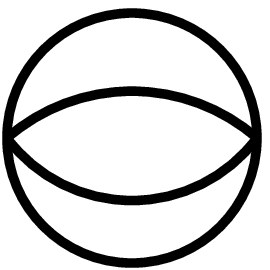,width=0.8cm}}
+\frac{1}{16} \; \; \raisebox{-2mm}{\epsfig{file=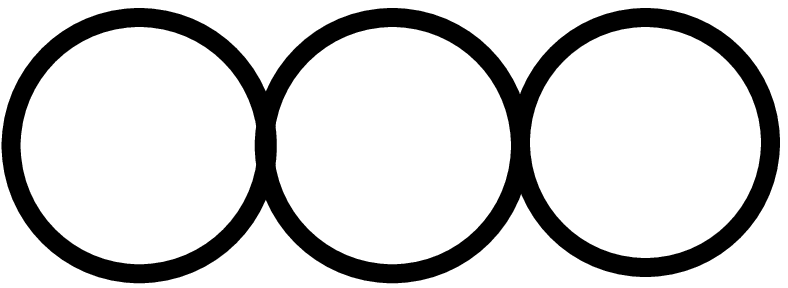,width=1.7cm}}
+\frac{1}{24} \; \; \raisebox{-3mm}{\epsfig{file=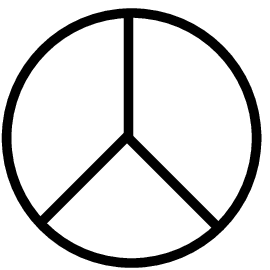,width=0.8cm}}
+\frac{1}{16} \; \; \raisebox{-2mm}{\epsfig{file=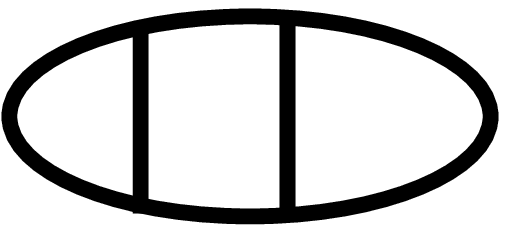,width=1.4cm}} \nn \\
&+&\frac{1}{8} \; \;  \raisebox{-3mm}{\epsfig{file=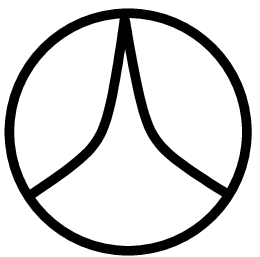,width=0.8cm}}
+\frac{1}{8} \; \;  \raisebox{-3mm}{\epsfig{file=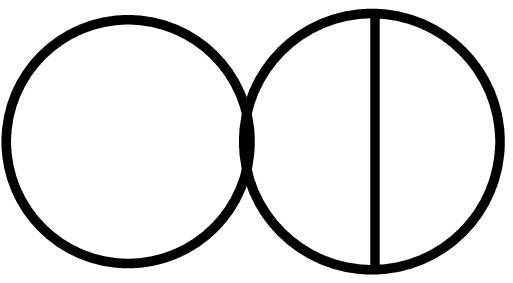,width=1.4cm}}
+\frac{1}{2} \left( \raisebox{-2mm}{\epsfig{file=2bubble.eps,width=1.4cm}}
   + \raisebox{-2mm}{\epsfig{file=sandwich.eps,width=0.7cm}} \right)^2 \nn \\
&=& \frac{1}{1152m^2x^2} (m-1) ( m^3+289m^2-1129m+1175 )
\eqa
The result for the asymptotic expansion of $H_m(x)$ to this order is:
\bq
H_m(x) \sim \frac{e^{mx}}{\sqrt{ m(2 \pi x)^{m-1}}} \{ 1 + \frac{m^2-1}{24mx}
+ \frac{1}{1152m^2x^2} (m-1) ( m^3+289m^2-1129m+1175 )+ \mathcal{O}(\frac{1}{x^3}) \}
\eq
and higher terms can be obtained in a similar way.

\vspace*{0.5cm}

\paragraph{Acknowledgments} We would like to thank Andr\'e van Hameren for lending 
an extra hand in the counting.

\end{document}